\documentclass[twocolumn, natbib209]{aastex62}
\usepackage{graphicx}
\usepackage{xspace}
\usepackage{natbib}
\usepackage{float}
\usepackage{epstopdf}
\usepackage{amsmath}
\usepackage{savesym}
\savesymbol{tablenum}
\usepackage{siunitx}
\restoresymbol{SIX}{tablenum}

\newcommand{\teff}{\ensuremath{T_{\rm eff}}}
\newcommand{\snr}{SNR$_{\lambda}$}
\newcommand{\M}{\ensuremath{M}}
\newcommand{\R}{\ensuremath{R}}
\newcommand{\Lum}{\ensuremath{L}}

\newcommand{\lamostMcatalog}{122,677\xspace} 
\newcommand{\lamostMcatalogunique}{103,467\xspace} 

\newcommand{\tcdlamostcross}{30,152\xspace} 
\newcommand{\tcdlamostcrossunique}{24,546\xspace}

\newcommand{\validationfinalcountunique}{11,279\xspace}
\newcommand{\validationfinalcount}{18,844\xspace}

\newcommand{\trainingfinalcount}{1,388\xspace} 

\newcommand{\testfinalcountdups}{33,095\xspace} 
\newcommand{\testfinalcount}{29,678\xspace} 

\newcommand{\rsun}{\ensuremath{R_\odot}}
\newcommand{\msun}{\ensuremath{M_\odot}}
\newcommand{\lsun}{\ensuremath{L_\odot}}

\newcommand{\teffcannonrange}{2901$<$\teff$<$4113~K}
\newcommand{\radiuscannonrange}{0.14$<\R<$0.66~\rsun}
\newcommand{\masscannonrange}{0.10$<\M<$0.71~\msun}
\newcommand{\lumcannonrange}{0.002$<\Lum<$0.115~\lsun}

\newcommand{\teffsig}{110\xspace K}
\newcommand{\radiussig}{0.065\xspace}
\newcommand{\masssig}{0.054\xspace}
\newcommand{\lumsig}{0.012\xspace}

\newcommand{\cannon}{\textit{The Cannon}\xspace} 
\DeclareMathOperator*{\argmax}{argmax}

\received{September 12, 2019}

\revised{December 9, 2019}

\accepted{December 11, 2019}

\submitjournal{AJ}

\shorttitle{Fundamental data-driven parameters of 30,000 M~dwarfs}
\shortauthors{Galgano et al. 2019}

\begin{document}

\title{Fundamental Parameters of $\sim$30,000 M~dwarfs in LAMOST DR1 \\ Using Data-driven Spectral Modeling}

\correspondingauthor{Brianna Galgano}
\email{brianna.galgano@gmail.com}

\author[0000-0001-5379-4295]{Brianna Galgano}
\affil{Fisk University, Department of Life and Physical Sciences, W.E.B. DuBois Hall, 1717 Jackson St, Nashville, TN 37208, USA}

\author[0000-0002-3481-9052]{Keivan Stassun} 
\affil{Vanderbilt University, Department of Physics \& Astronomy, 6301 Stevenson Center Lane, Nashville, TN 37235, USA}

\author[0000-0002-0149-1302]{B\'arbara Rojas-Ayala}
\affil{Instituto de Alta Investigaci\'on, Universidad de Tarapac\'a, Casilla 7D, Arica, Chile}

\begin{abstract}
M~dwarfs are the most common type of star in the Galaxy, and because of their small size are favored targets for searches of Earth-sized transiting exoplanets. Current and upcoming all-sky spectroscopic surveys, such as the Large Sky Area Multi-Object Fiber Spectroscopic Telescope (LAMOST), offer an opportunity to systematically determine physical properties of many more M~dwarfs than has been previously possible. Here we present new effective temperatures, radii, masses, and luminosities for \testfinalcount M~dwarfs with spectral types M0---M6 in the first data release (DR1) of LAMOST. We derived these parameters from the supervised machine learning code, \cannon, 
trained with \trainingfinalcount M~dwarfs in the Transiting Exoplanet Survey Satellite (TESS) Cool Dwarf Catalog that were also present in LAMOST with high signal-to-noise ratio ($>$250) spectra. Our validation tests 
show that the output parameter uncertainties are strongly correlated 
with the signal-to-noise of the LAMOST spectra, and 
we achieve typical uncertainties of \teffsig\ in \teff\ ($\sim$3\%), \radiussig~\rsun\ ($\sim$14\%) in radius, \masssig~\msun\ ($\sim$12\%) in mass, and \lumsig~\lsun\ ($\sim$20\%) in luminosity.
The model presented here can be rapidly applied to future LAMOST data releases, significantly extending the samples of well characterized M~dwarfs across the sky using new and exclusively data-based modeling methods.
\end{abstract}


\section{Introduction} \label{sec:INTRODUCTION}

M~dwarfs are preferred targets for exoplanet hunting due to their status as the most common stellar type in the galaxy. \citep[e.g.,][]{Mann:2015}, and for their potential in detecting Earth and super-Earth sized exoplanets using both ground and space-based telescopes via radial velocity measurements and transit detection \citep[e.g.][]{Shields:2016}. The fundamental properties of any host star such as radius and mass need to be known to a reasonable degree of certainty in order to characterize the exoplanets they host. 

For nearby, bright, well-known M~dwarf samples identified via, e.g., very high proper motions \citep[e.g.,][]{Lepine:2013}, precise stellar properties can be estimated from broadband colors and parallax alone. For example, \citet[e.g.,][]{Mann:2019, Mann:2015} have developed empirical calibrations that permit the determination of radii and masses of M-dwarfs to better than 10\% precision from near-infrared colors and a near-infrared absolute magnitude. However, for the multitude of M~dwarfs that have not yet been identified or characterized by such catalogs, there remains a need to develop robust spectroscopic methods which can circumvent problems associated with reddening that can negatively affect color-based methods, and additionally leverage current and upcoming large-scale spectroscopic surveys. 

However, automated M dwarf characterization via stellar spectroscopy with synthetic models is difficult given the complicated nature of their stellar atmospheres. For example, generating model synthetic spectra for stars cooler than FGK-types, such as M~dwarfs, is difficult because of the formation of molecules in their photospheres e.g., TiO, VO, and CaH in the optical, and H$_2$O and CO in the near-IR \citep[ e.g.,][]{Rojas-Ayala:2011, Shields:2016}. Very heavy absorption can be observed in an M~dwarf spectrum by the presence of molecular compounds which are allowed under their lower effective temperatures (\teff\ $<$ 3850~K), which lead to broad and overlapping spectral lines that can be hard to parse \citep[ e.g.,][]{Rojas-Ayala:2011}. Modeling the absorption lines of these compounds is incomplete \citep[e.g.,][]{Allard:2011, Husser:2013}, and the lack of a well-defined continuum is also expected, with some wavelength regions being completely saturated with absorption features, making normalization for standard spectroscopic analysis also difficult. Additional complications due to the stars' intrinsic properties are deep convection cells and magnetic activity/rapid rotation, effects that in general are not included in models of low-mass stellar atmospheres. 

The low luminosities of these stars also brings restrictions to both of these methods; parallax of dwarfs can only be obtained for the very brightest, and quality (high signal-to-noise ratio) spectra are difficult to obtain. 
A common method of practice is to also parameterize the properties of dwarfs if they are in binary with FGK-type companions, but these are limited to a select number of dwarfs.

The result is that physics-based models are incomplete and computationally expensive for M~dwarfs which make the method of fitting to synthetic spectra difficult. This presents a problem---there is a high demand for M~dwarf basic properties for exoplanet and stellar population characterization, but there is much room for improvement in automated analysis in terms of both computational speed and accuracy.

A promising alternative is using data-based modeling approaches that use machine learning to predict what a spectrum would look like given its stellar properties, and then infer those properties of uncharacterized spectra based on the model. We discuss in this work how \cannon\footnote{\url{https://github.com/annayqho/TheCannon}} \citep{Ness:2015} can successfully model low resolution, low to moderate signal-to-noise M~dwarf spectra, and we use \cannon model to find the properties of \testfinalcount previously uncharacterized optical spectra of M~dwarfs in the LAMOST DR1 catalog. 

\cannon has so far been successfully applied to characterize red giants using APOGEE spectra and LAMOST spectra and using parameters derived via the \textit{ASPCAP} pipeline \citep{Ho:2017_lamostgiants, Ho:2017_labeltransfer}. \cannon has also been applied to much higher resolution than LAMOST M~dwarf spectra \citep{Behmard:2019}.
LAMOST is an optical spectroscopic survey all-sky survey for the Northern celestial hemisphere \citep{Luo:2015}. There are $\sim$121,000 M~dwarf spectra in the first data release (DR1), the majority of which have not yet been characterized beyond a simple spectral subtype determined from color photometry. We cross-match the LAMOST targets with the TESS Cool Dwarf Catalog \citep{Muirhead:2018}, from which we create a training set to train a spectral model using \cannon. 

In Section~\ref{sec:DATA}, we summarize the data selection and preparation in this work. In Section~\ref{sec:METHODS}, we discuss the methods by which we chose an optimal training set for accuracy and the testing set for applicability. We also discuss the methods by which we assessed the model's accuracy in determining basic parameters: \teff, radius, mass, and luminosity. In Section~\ref{sec:RESULTS}, we report these properties and their uncertainties for the \testfinalcount M~dwarfs in LAMOST DR1 for which we were able to determine reliable properties. Finally, Section~\ref{sec:SUMMARY} concludes with a summary of our conclusions. 

\section{Data} \label{sec:DATA}

In this section, we describe the data that we use, their preparation for use with \cannon, and quality control steps taken to train, validate, and apply \cannon model to reliable data. 

The data are drawn from two main catalogs: (1) the LAMOST first data release (DR1) M~dwarf catalog, is the spectral data set that we wish to classify \citep{Luo:2015}, and (2) the TESS Cool Dwarf Catalog (TCD) \citep{Muirhead:2018}, whose parameters we utilize for training and validation of \cannon model. 
The parameters from the TCD \citep{Muirhead:2018} on which we have chosen to train \cannon model are effective temperature (\teff), radius (\R), mass (\M), and luminosity (\Lum); these are the parameters that we will therefore be able to estimate for the LAMOST DR1 stars. 

Our data are split into three subsets, following the approach described by \citet{Ness:2015}. The subset of \tcdlamostcross\ TCD stars for which we also have LAMOST spectra (hereafter, the cross-matched TCD) are divided into a \textit{training set} and a \textit{validation set}. The training set contains the highest signal-to-noise ratio (SNR) spectra of all TCD and LAMOST cross-matches, paired with their known associated stellar characteristics or parameters \citep[i.e., \textit{labels} in][]{Ness:2015}; the training set is used to generate the data-driven model. The validation set is the remaining cross-matched TCD stars that were not used in the training set but whose parameters are known and therefore can be used to assess the fidelity of the model. The LAMOST stars that are included in neither the training set nor the validation set comprise the \textit{survey set}. These are spectra we wish to characterize using the model we created from the training set, and vetted with the validation set. 

\subsection{LAMOST DR1 Spectra} \label{subsec:lamost_spec}

The Large Sky Area Multi-Object Fiber Spectroscopic Telescope (LAMOST) is a low resolving power (R $\sim$ 1800), optical/near-infrared (3690--9100\AA), ground-based survey \citep{Luo:2015}. One of the data products from LAMOST is the LAMOST DR1 M~dwarf catalog \citep{Guo:2015}, a catalog of stars whose spectra have been classified broadly as being of M spectral type but without more detailed physical information. This catalog consists of \lamostMcatalog\ spectra with simple spectral type classifications from M0 to M9; these spectra represent \lamostMcatalogunique\ unique objects after excluding any duplicate observations of the same target.

\begin{figure}[!ht]
\includegraphics[width=\linewidth]{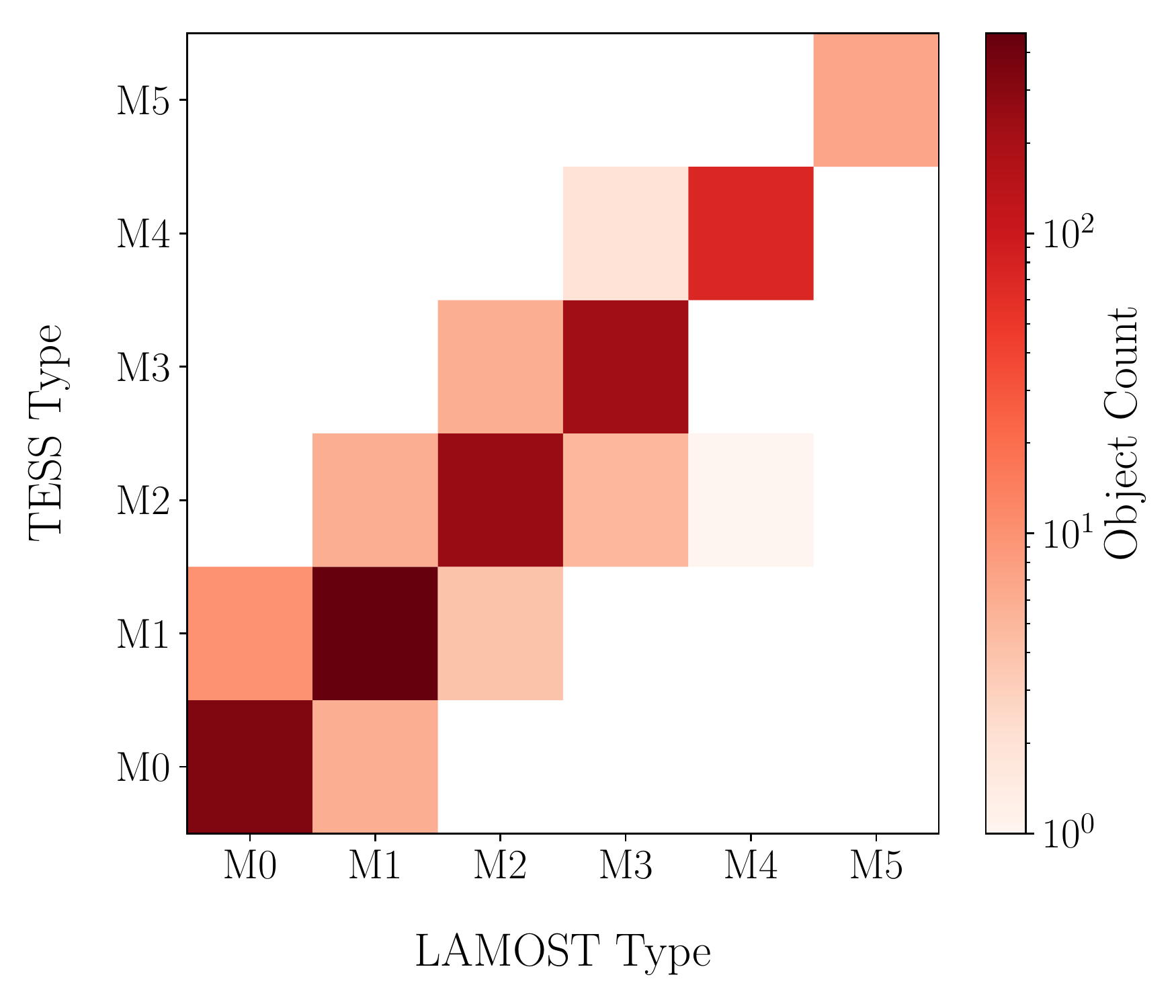}
\caption{The spectral subtype agreement between cross-matched M stars of LAMOST DR1 and the TESS Cool Dwarf Catalog (TCD). 
Note the lack of cross-matched stars with spectral types later than M6; this effectively sets the cool limit of our analysis.
\label{fig:spectype_tessxlamost_1to1}}
\end{figure}

While the LAMOST catalog includes some stars with approximate classifications as late as M9, we chose to include only stars with a LAMOST DR1 spectral subtype of M6 and earlier, given that the training set includes only stars with LAMOST subtypes from M0 to M6 (see Section~\ref{subsec:model_training}). Figure~\ref{fig:spectype_tessxlamost_1to1} shows the comparison of the LAMOST spectral subtypes to those from the TCD for the \tcdlamostcross spectra that were cross-matched, where the relatively small number of objects with the latest spectral types is due to difficulty of observing very faint objects with high signal.

The LAMOST DR1 catalog provides the spectra for each of these stars. We require that to be usable a LAMOST spectrum must include a wavelength solution ($\lambda$), flux ($f$), and flux error (or flux inverse variance; $1/\sigma_{f}^2$), the last of which is used as a statistical weight for training a model pixel-by-pixel. 

\subsubsection{Basic processing of LAMOST spectra}\label{subsubsec:lamost_spec_processing}

First, we applied a bad pixel mask using the code supplied with \cannon\ package for this purpose. The bad pixel mask flagged pixels in the spectra from information provided in the LAMOST meta-data, including poor sky-subtraction, bad CCD pixels, and infinite or negative flux inverse variance values. The mask also manually flagged pixels with strong telluric lines and discontinuities from the joining of LAMOST's blue and red spectrographs at 5800--6000 \AA. For pixels flagged in the bad pixel mask we manually set the inverse variances to a very small value\footnote{We chose a value of $10^{-5}$, as opposed to 0, which helps avoid errors arising from division by zero in future operations.}.

In order for a spectrum to be input into \cannon for analysis, the spectra in the training, validation, and survey subsets also must be interpolated to a common wavelength scale with uniform wavelength spacing. From inspection of the spectra we found that the LAMOST data were most consistently free of bad pixels or other problems at wavelengths shorter than $\sim$4500\AA, and thus we interpolated to a common wavelength range of 4500--7500\AA, and used a uniform spacing of 1~\AA, as these parameters approximately represent the native format of the LAMOST spectra for M stars. Each resulting spectrum is a $3000 \times 3$ data array, with each of the 3000 pixels containing a standardized wavelength, flux, and a flux inverse variance float value. 

\begin{figure}[!ht]
\includegraphics[width=\linewidth]{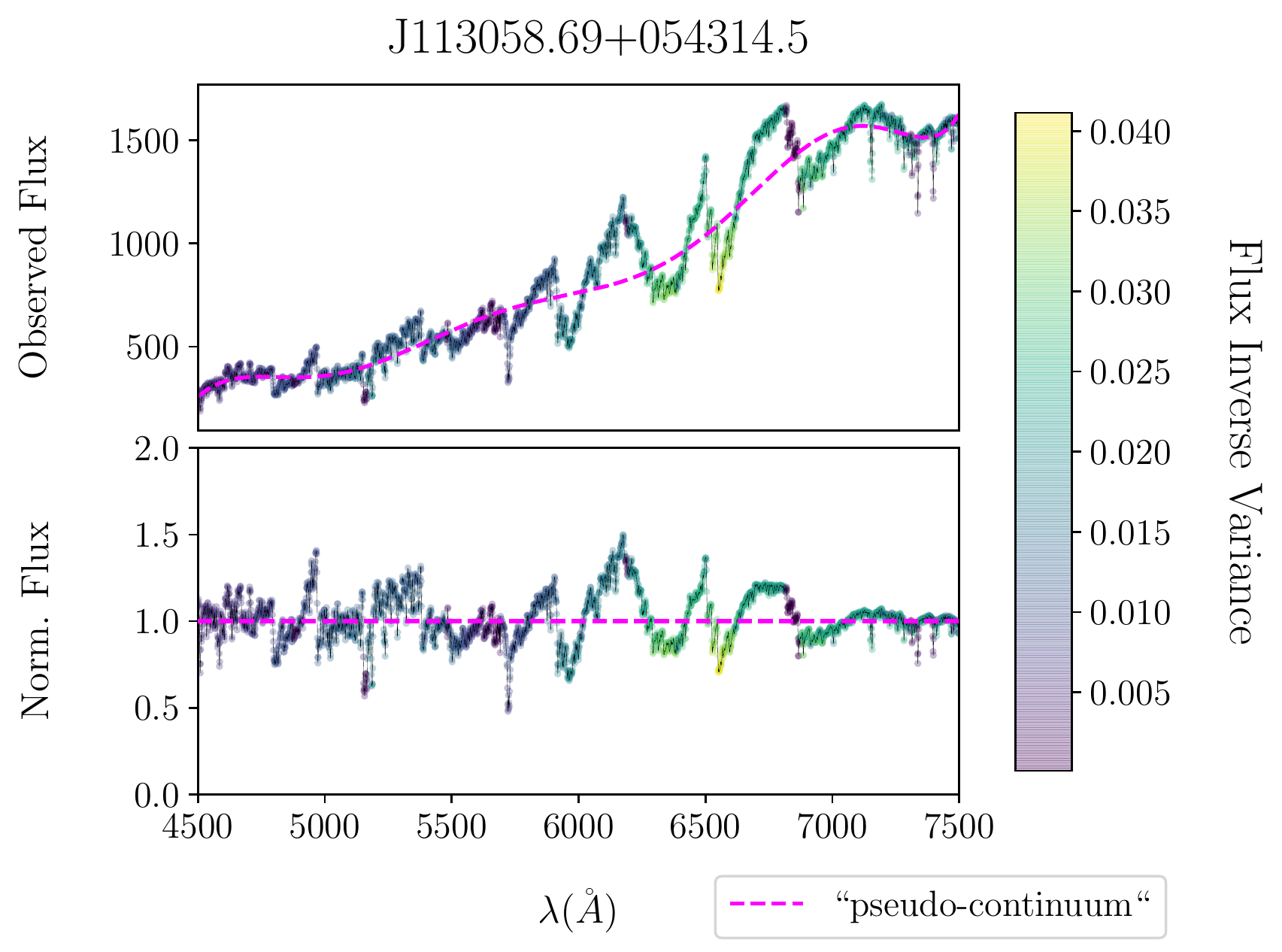}
\caption{An example of a pseudo continuum-normalized using Gaussian smoothing LAMOST spectrum of an M~dwarf from our training set. The effective Gaussian smoothing width is 50~\AA. All spectra from each data subset (training, validation, and survey) are normalized with this continuum fitting method as defined by \cite{Ness:2015} (see Section \ref{subsec:lamost_spec}).}
\label{fig:normalized_spectrum}
\end{figure}

In order to compare fluxes from spectrum to spectrum to create \cannon model, each spectrum's continuum must also be normalized. Traditional methods of finding and then dividing an M~dwarf's spectrum by its continuum (continuum-normalization) are challenging because of the complex nature of the stellar spectra as mentioned in Section \ref{sec:INTRODUCTION}. Therefore, we adopted the so-called ``pseudo continuum-normalization" method used by \citet{Ho:2017_labeltransfer,Behmard:2019}, which first approximates the continuum at wavelength $\lambda_0$ via Gaussian smoothing over $L$ pixels and then normalizes the original spectrum with this smoothed fit: 
\begin{equation}
\label{eq:norm}
f(\lambda_0) = \frac{\sum_n f_n{\sigma_{n}}^{-2}w_n(\lambda_0)}{ \sum_n {\sigma_{n}}^{-2}w_n(\lambda_0)}
\end{equation}
where index $\sum_n$ represents the sum over pixels, and the individual pixel weights, $w_n$, are given by: \begin{equation}
\label{eq:norm_weight}
w_n(\lambda_0) = e^{-\frac{(\lambda_{0}-\lambda_{n})^2}{L^2}}
\end{equation}
where $\lambda_{n}$, $f_{n}$,  $\sigma_{n}$, is the wavelength, flux, and flux error at the $n$th pixel, respectively. We pseudo-normalized all spectra with a Gaussian width of $L$ = 50~\AA~pixels, which is wider than most broader absorption features, and used the flux inverse variance as weights as recommended by \citet{Ho:2017_labeltransfer}. The procedure is visualized with an example in Figure~\ref{fig:normalized_spectrum}. 

We note that this is not a standard robust normalization, in which the maximum flux is well defined and set to unity, but it is a technique where the spectra can be quantitatively and consistently compared because they are placed on a common flux scale. This pseudo-normalization method sometimes over-fits too deeply into absorption bands (e.g., Figure~\ref{fig:normalized_spectrum}), but because the LAMOST spectra have the same Gaussian function outlined in \cite{Ho:2017_labeltransfer} systematically applied, it is the preferable technique as it avoids human or physics-model bias from other standard normalization methods.

Next, LAMOST DR1 objects with no redshift $Z$ parameter reported in the meta-data cannot be shifted to a rest wavelength frame, a requirement for any spectrum to be input into \cannon. Approximately 10,000 stars were eliminated for this reason. 

Finally, through visual inspection of a random set of the LAMOST spectra for quality checks, some were found to have unusually high flux at blue wavelengths ($\sim$4500\AA\ and bluer) which is atypical of M~dwarf stars and almost certainly not intrinsic. We chose to eliminate $\sim$1000 stars based on having especially severe upward trends to the blue, which we defined on the basis of the slope of the pseudo-continuum being more negative than $-3$ (in units of normalized flux per unit wavelength). The cause of this high blueward flux effect is unclear, but might be due to instrumental, calibration, or sky subtraction error. 

\subsubsection{Spectrum filtering quality control}\label{subsubsec:spectrum_filtering}

We require an objective measure of the quality of the LAMOST spectra, both to select the highest quality subset for training, and to ensure that we do not attempt to classify very poor spectra. 

We chose as our quality metric a measure of signal-to-noise ratio (SNR), which we calculated from the sum of the measured SNRs in the individual LAMOST quasi-photometric bands ($ugriz$), hereafter \snr. This approach was preferred over the more formal method of using flux error as it is computationally faster and independent of the bad pixel masks, and it is readily obtainable from the LAMOST meta-data. We determined a requirement of \snr\ $>$ 50 based on validation testing of the model described in Section~\ref{subsec:error_estimation}; consequently $\sim$45,000 objects with \snr\ $<$ 50 were eliminated from any further analysis.

\begin{figure}[!ht]
\centering
\includegraphics[scale=0.49]{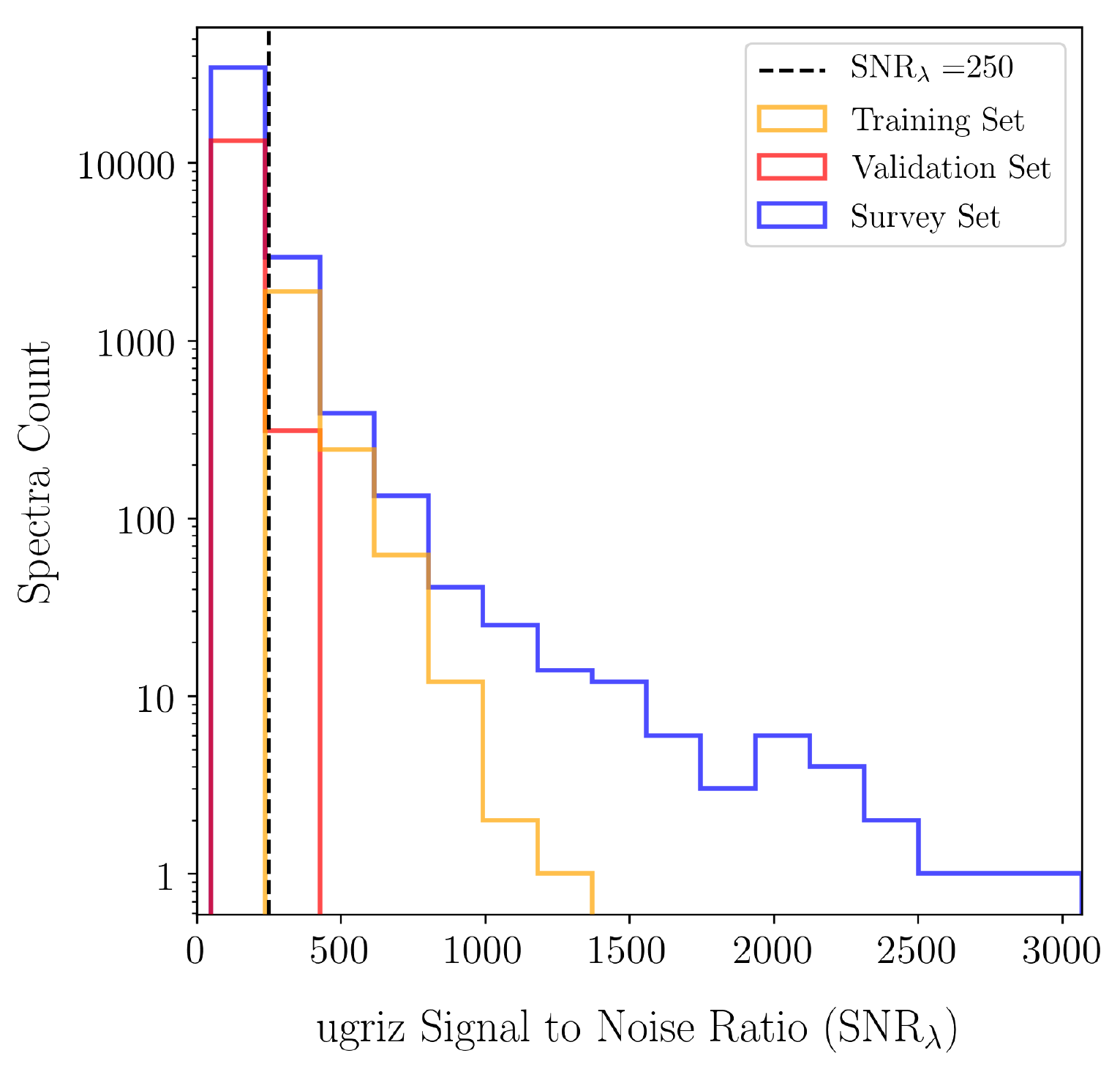}
\caption{Signal to noise ratio measured in $ugriz$ bands (\snr) for both survey and training sets. Cutoff minimums chosen were \snr $>$250 for the training subset and $>$50 for the validation and survey subset (see Section \ref{subsec:snr_quality_cutoffs}). \label{fig:snrs_test_training_sets}}
\end{figure}

For the training subset, we chose to use a more stringent quality threshold of \snr\ $>$ 250 (see Figure~\ref{fig:snrs_test_training_sets}). These choices of \snr\ thresholds are described more fully in Section~\ref{subsec:model_validation}. 

\subsection{TESS Cool Dwarf Parameters for Training and Validation} \label{subsec:get_TESS_parameters} 

The training and validation of \cannon model requires a set of stars whose physical parameters \citep[i.e., {\it labels} in][]{Ness:2015} are known. We adopt the parameters provided in the TCD as the known labels with which to train and validate our model. Those parameters were in turn determined by \citet{Muirhead:2018} on the basis of the empirical relations of \citet{Mann:2015}.

Note that while the TCD was created in order to enhance planet detection around cool dwarfs by the {\it TESS} mission, the TCD was created more generally to develop a systematically vetted, brightness and proper-motion limited sample of cool dwarfs across the sky in a manner that is as unbiased as possible. We cross-matched the LAMOST DR1 M-dwarf catalog with the entirety of the $\sim$1~million sources in the TCD catalog, not just the TCD targets that were ultimately selected by the {\it TESS} mission for 2-minute cadence observations for planet detection. Of course, our training set will necessarily inherit any features inherent in the TCD sample, the most important of which are as follows \citep[see][]{Muirhead:2018}: Unresolved binaries will appear to have masses and radii that are too large for their color; metallicities are not known for the majority of the TCD sample, hence we cannot train on metallicity as a data label and we cannot incorporate any metallicity effects into our trained model; the catalogs from which the TCD drew its proper motions were highly incomplete in the Southern hemisphere; and some stars in the TCD relied on inhomogeneous sources of photometry for colors.

Upon a quality check examination of \teff\ vs.\ $M$ and \teff\ vs.\ $R$ for these stars, we found a small number of stars whose TCD parameters deviate from the \citet{Mann:2015} empirical relations. These outliers are most likely the result of inhomogeneous sources of photometry for some stars and/or unresolved binaries, as noted above.

In order to ensure consistency in the parameters used for the training, we eliminated these 18 outlier stars that deviated in \R\ by more than 0.01~\rsun\ and in $M$ by more than 0.01~\msun.

\section{Methods}\label{sec:METHODS}

In this section we discuss the methods for ensuring the accuracy of the model produced by \cannon for M~dwarfs and detail how we further optimized the data-driven model by choosing appropriate \snr\ minimum cutoffs for training, validation, and survey sets.

Parameter uncertainties are currently not incorporated in \cannon's generative modeling; broadly, the model output uncertainties are formal uncertainties based on the flux error and closeness of the model fit to a spectrum. We therefore determine more robust uncertainties on the output parameters using the scatter of the validation set about known parameters for the \tcdlamostcross TCD/LAMOST cross-matches, to find a function that estimates parameter error with \snr, as described in Section~\ref{subsec:error_estimation}.

\subsection{Model Training}\label{subsec:model_training}

In the first step, we trained \cannon with the training set of \trainingfinalcount stars for which we have high-quality LAMOST spectra as well as parameters (\teff, \R, \M, and \Lum) from the TCD (see Section~\ref{subsec:get_TESS_parameters}), and that satisfied the quality control cuts discussed in Section~\ref{sec:DATA}. 

As outlined in previous work \citep[e.g.,][]{Ness:2015, Ho:2017_labeltransfer, Behmard:2019}, we can represent a given spectrum's flux at wavelength $\lambda$, $f_\lambda$, as the product of the model coefficients $\theta_\lambda$ and the spectrum's labels (i.e., physical parameters), $l$, plus noise: 
\begin{equation}  
f_{\lambda} = \theta_\lambda \cdot l + noise
\end{equation}

The noise is the quadrature sum of the measurement error in the observed flux, $\sigma_{\lambda}$, and the uncertainty in the model coefficients, $s_{\lambda}$: 

\begin{equation}  
noise = \lbrack s_{\lambda}^2+\sigma_{\lambda}^2\rbrack \zeta_{\lambda}
\end{equation}

where $\zeta_{\lambda}$ is a random Gaussian deviate with zero mean and zero unit variance \citep{Ness:2015}. For our implementation with training labels taken from the TCD, we define $l$ in terms of the physical parameters \teff, $R$, $M$, and \Lum, to second order and including cross-terms:

\begin{equation}         
\begin{aligned}
{l} \equiv & \lbrack 1, \teff, R, M, L, \teff\cdot R, \teff\cdot M, \teff\cdot L, R\cdot M, \\ & R\cdot L, M\cdot L, \teff^{2}, R^{2}, M^{2}, L^{2} \rbrack \\
\end{aligned}
\end{equation}

where the first term allows for a linear offset in the fitted flux values \citep{Behmard:2019}.

For a given spectrum, $n$, we can then get the single-pixel log likelihood function ($\ln p$), which gives the most likely flux value at a specific wavelength pixel, given model scatter, observational uncertainty, and set of labels:

\begin{equation} 
\begin{aligned}
    \ln p (f_{n\lambda} | \theta_\lambda^T, l_n, s_{\lambda}^2) = \frac{1}{2} \frac{\lbrack f_{n\lambda}-\theta_\lambda^T \cdot l_n \rbrack^2}{s^2_{\lambda} + \sigma^2_{n\lambda}} - \frac{1}{2}\ln (s^2_{\lambda} + \sigma^2_{n\lambda})
\end{aligned}
\end{equation}

where the superscript $T$ denotes the transpose of the matrix. 
\cannon then uses these pixel likelihood probabilities to derive model coefficients that apply to the full set of \citet[e.g.,][]{Ness:2015}:

\begin{equation}
\theta_{\lambda}, s_{\lambda} \leftarrow \argmax_{\theta_{\lambda}, s_{\lambda}} \sum_{n=1}^{N} \ln p (f_{n\lambda} | \theta_\lambda^T, l_n, s_{\lambda}^2)
\end{equation}

The output of this step is a set of model coefficients for each wavelength pixel for each stellar parameter. The data-driven model can now be applied to any other LAMOST spectrum, where the most probable value for a stellar parameter can be calculated given flux and flux error across all pixels.

\subsection{Model Validation}\label{subsec:model_validation}

\begin{figure*}[!ht]
\plotone{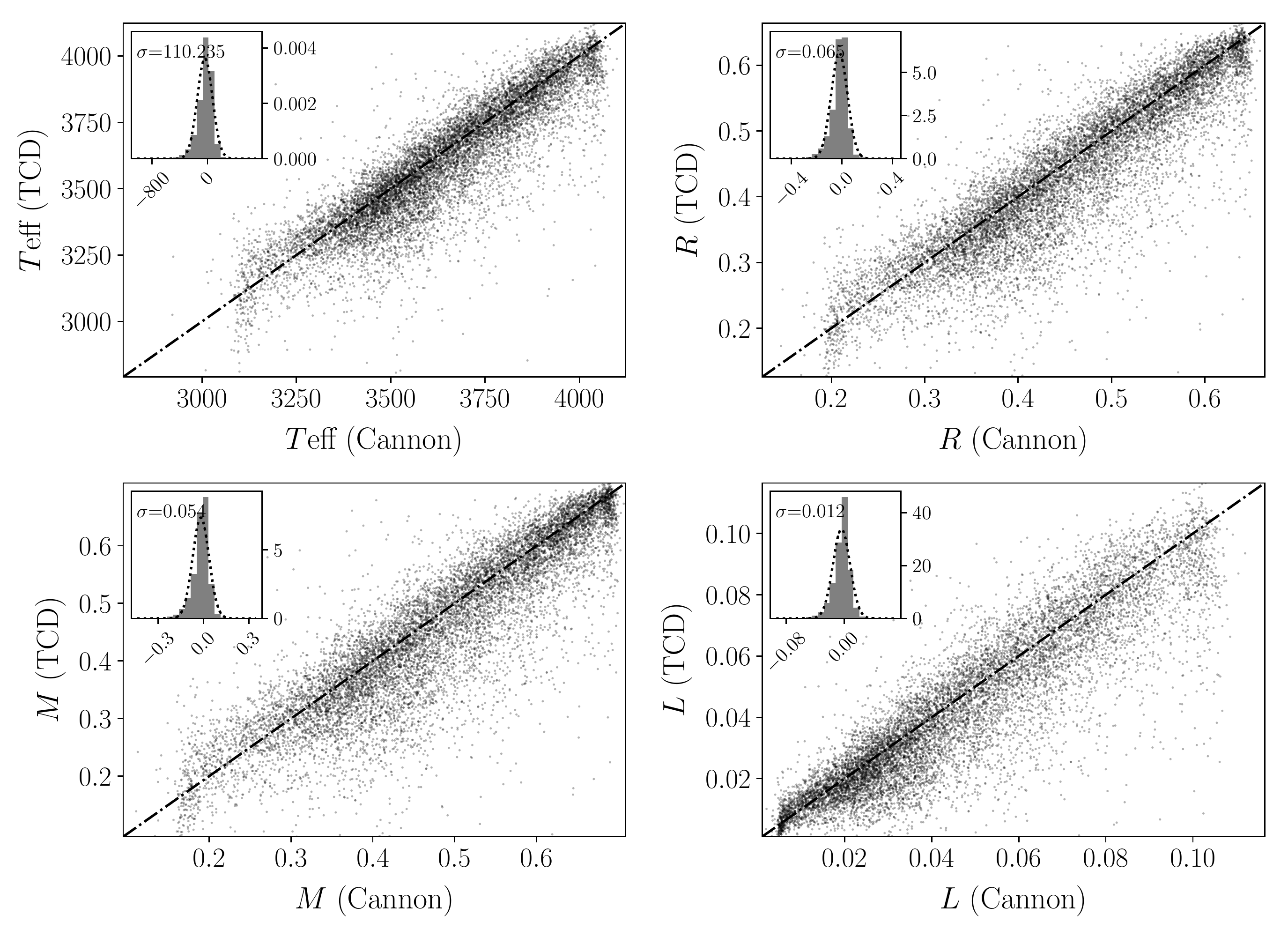}
\caption{Comparison of output to input labels for the validation data set, cross-matched between LAMOST DR1 and TCD. While there is a general one-to-one relationship between the model-derived values and TCD, the model performs worse at either end of the range of M subtypes (very early or late) as discussed in Section~\ref{subsec:model_validation}. A total of \validationfinalcount spectra, or \validationfinalcountunique individual LAMOST objects comprises the validation set.}\label{fig:tessxlamost_1to1}
\end{figure*}

Next, to validate \cannon model, we applied the trained data-driven model from above to the \tcdlamostcross\ stars comprising the validation subset (see Section~\ref{sec:DATA}). We chose \validationfinalcount LAMOST spectra after the same quality checks described in \ref{sec:DATA}, except the \snr\ minimum is reduced to $>$50. Figure~\ref{fig:tessxlamost_1to1} shows the results of the validation, in which we compare the parameters output by \cannon to the input values as adopted from the TCD.

Overall, we observe good one-to-one agreement between the TCD input and \cannon output labels for the validation set in all four of the training labels (\teff, \R, \M, and \Lum). The scatter about the one-to-one line of agreement is approximately Gaussian (see the insets in each panel of Figure~\ref{fig:tessxlamost_1to1}), with Gaussian widths of $\sim$100~K, $\approx$0.06~$R_\odot$, $\approx$0.05~$M_\odot$, and $\approx$0.01~$L_\odot$ in \teff, \R, \M, and \Lum, respectively. For reference, the uncertainties in the training labels themselves as reported in the TCD are $\approx$100~K, $\approx$0.06~$R_\odot$, $\approx$0.07~$M_\odot$, and $\approx$0.008~$L_\odot$, respectively in \citet{Muirhead:2018}.

We do observe some slight asymmetry in the residuals about the one-to-one relation, or a ``tail" of residuals where \cannon output labels are a bit larger than the analogous values of TCD input labels. We also observe some evidence for breakdown of the model for the coolest stars, i.e., \teff$\lesssim$3150~K (likely due to the small number of M5--M6 stars in the training set), as well as for the hottest stars, i.e., \teff$\gtrsim$4000~K, where the scatter in the residuals become slightly discontinuous and asymmetric (Figure~\ref{fig:tessxlamost_1to1}). The disagreement at the hotter end ($\gtrsim$4000~K) is not surprising, as the TCD's upper \teff\ limit is $\sim$4000~K \citep{Muirhead:2018}.

\subsection{\snr\ Quality Cutoffs}\label{subsec:snr_quality_cutoffs}

Several factors such as wavelength range, quantity of objects in the training set, type of continuum-normalization, label choice, and others, can affect the accuracy of the labels derived by our trained model from \cannon package. In our tests with \cannon model applied to the LAMOST spectra, we found that SNR of spectra, both for training and validation, was the most important factor for model accuracy.

\begin{figure*}[!ht]
\centering 
\includegraphics[scale=0.57]{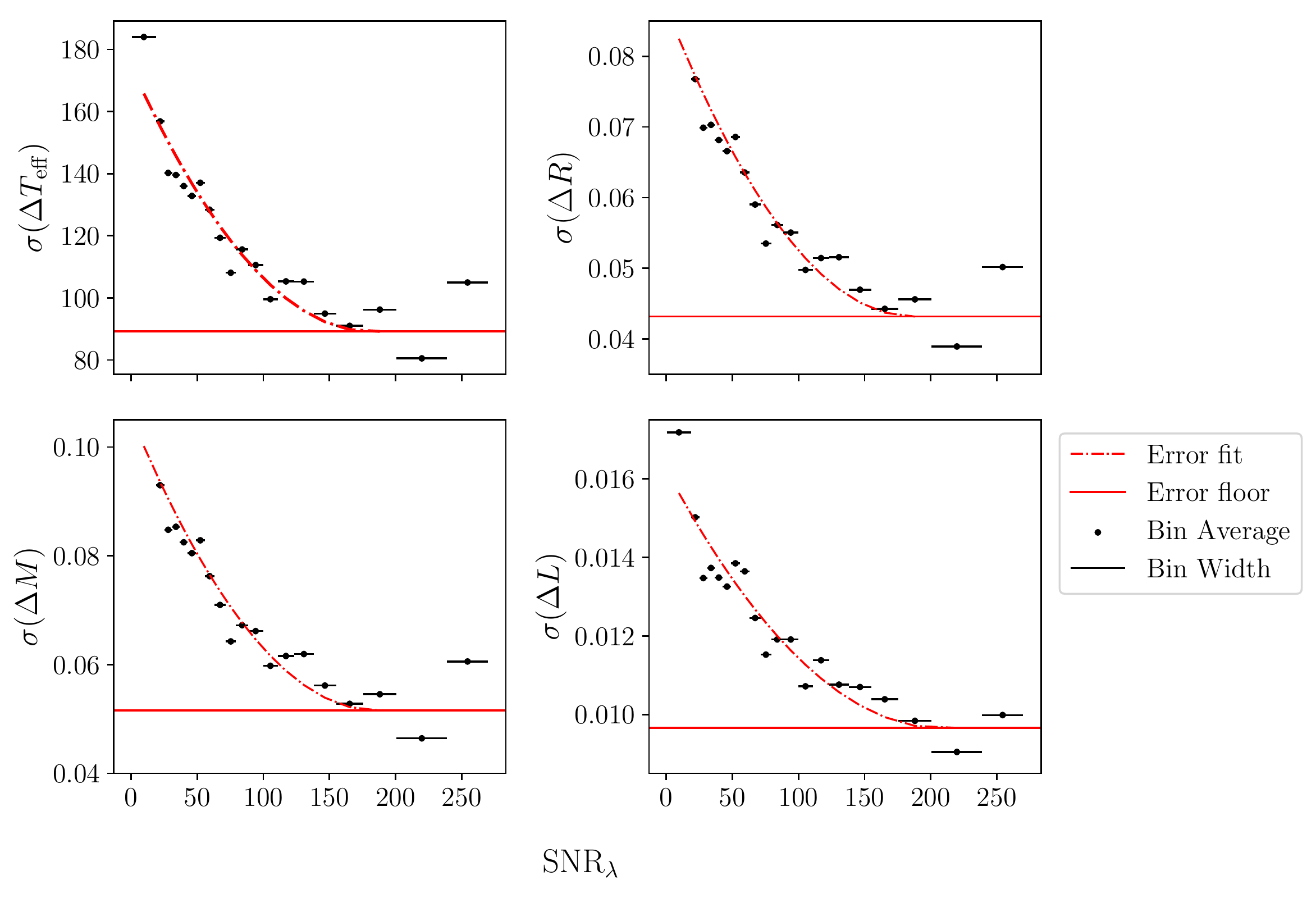}
\caption{Relationship between \snr\ and label residuals between cross-matches of LAMOST DR1 and TCD. Each point represents the bin average and its horizontal bar represents that bin's width. We note bin width increases at high \snr\ so that each bin contains approximately the same number of objects (1360$\pm$5; 860 in the bin for \snr\ $>$ 239). A small number of stars beyond \snr\ $>$ 300 are not shown. The coefficients of the best-fit 2nd degree polynomial are provided in Table~\ref{tab:error_coefficents}. The solid line represents the adopted error floor for \snr\ $>$ 225. 
\label{fig:snr_function_error_estimate}}
\end{figure*}

In order to determine an appropriate requirement for \snr\ for both the final training subset described above and for application of the trained model to the final survey set described below, we examined \cannon's performance as a function of \snr\ in Figure~\ref{fig:snr_function_error_estimate}.

We found that the accuracy of the model worsened significantly at \snr\ $<$~50, and so have opted in the final results reported below to exclude such stars. For example, at \snr $\lesssim 50$, the scatter in the inferred radii and masses becomes $\gtrsim$15\% (see Figure~\ref{fig:snr_function_error_estimate}). In addition, we selected a threshold of \snr\ $>$~250 for the training set, as these represent the very highest quality spectra while still possessing a significant number of objects ($\sim$2300) to include in the training set. Moreover, inspection of the trends in Figure~\ref{fig:snr_function_error_estimate} reveal that the scatter in the inferred parameters does not improve beyond \snr $\approx 250$, thus we assume that including all objects with \snr $>250$ in the training set will maximize the performance of the trained model. With this cutoff of \snr $< 250$ and other quality checks, the final training set consists of \trainingfinalcount spectra that are each unique objects (with unique LAMOST designations) out all \tcdlamostcrossunique unique objects available that are cross-matches with TCD.

\section{Results}\label{sec:RESULTS}

After application of the quality and other cuts described in the preceding sections, we have been able to successfully apply \cannon to \testfinalcount M~dwarfs, for which we report newly characterized \teff, \R, \M, and \Lum\ labels. A sample of these final results are reported in Table~\ref{tab:final_test_results}.

\begin{deluxetable*}{llllllllllll}[!ht]
\tablecolumns{12}
\tabletypesize{\small}
\tablehead{\colhead{Designation} & \colhead{\teff} & \colhead{$\sigma$\teff} & \colhead{\R} & \colhead{$\sigma$\R } & \colhead{\M} & \colhead{$\sigma$\M} & \colhead{\Lum} & \colhead{$\sigma$\Lum} & \colhead{\snr} & \colhead{SpType} & \colhead{$\chi^{2}$}}
\tablewidth{\linewidth}
\startdata
 J080546.84+202017.8 &   {3469.6} &    {128.5} &        {0.376} &        {0.064} &      {0.384} &      {0.077} &     {0.026} &     {0.013} &   {58.4} &          M4 &    {1998.0} \\
 J064420.73+323838.6 &   {3680.1} &    {106.8} &        {0.481} &        {0.054} &      {0.513} &      {0.063} &     {0.049} &     {0.014} &  {264.2} &          M1 &    {1899.0} \\
 J131151.74+565947.5 &   {3837.4} &     {99.0} &        {0.560} &        {0.049} &      {0.607} &      {0.059} &     {0.070} &     {0.012} &  {119.9} &          M1 &    {1312.0} \\
 J095520.46+301012.0 &   {3732.4} &    {129.2} &        {0.513} &        {0.064} &      {0.551} &      {0.077} &     {0.055} &     {0.013} &   {57.3} &          M2 &    {3592.7} \\
 J100744.45$-$020217.4 &   {3995.9} &    {124.2} &        {0.626} &        {0.062} &      {0.677} &      {0.074} &     {0.095} &     {0.013} &   {65.2} &          M0 &    {2512.6} \\
\enddata
\caption{Resulting stellar parameter values from applying \cannon M~dwarf model to LAMOST spectra. The formal error (1$\sigma$) has been estimated from the best fitted polynomial function of \snr\ (see Section~\ref{subsec:error_estimation}, Figure~\ref{fig:snr_function_error_estimate}). 
Units for \teff, \R, \M, and \Lum\ are K, $R_{\odot}$, $M_{\odot}$, and $L_{\odot}$, respectively. Spectral types are as reported by LAMOST DR1. The full table is available in the electronic version of the Journal; a small sample is provided here for guidance regarding its form and content.
\label{tab:final_test_results}}
\end{deluxetable*}

\subsection{Final Eliminations} \label{subsec:final_eliminations}

We note that a small fraction of the stars analyzed and that otherwise satisfied our quality cuts nonetheless failed to be characterized by \cannon. 
This was the case for 2,445 stars, for which at least one or more of the output labels failed, identified by {\tt NaN} or negative label values.

\begin{figure}[!ht]
\includegraphics[scale=0.49]{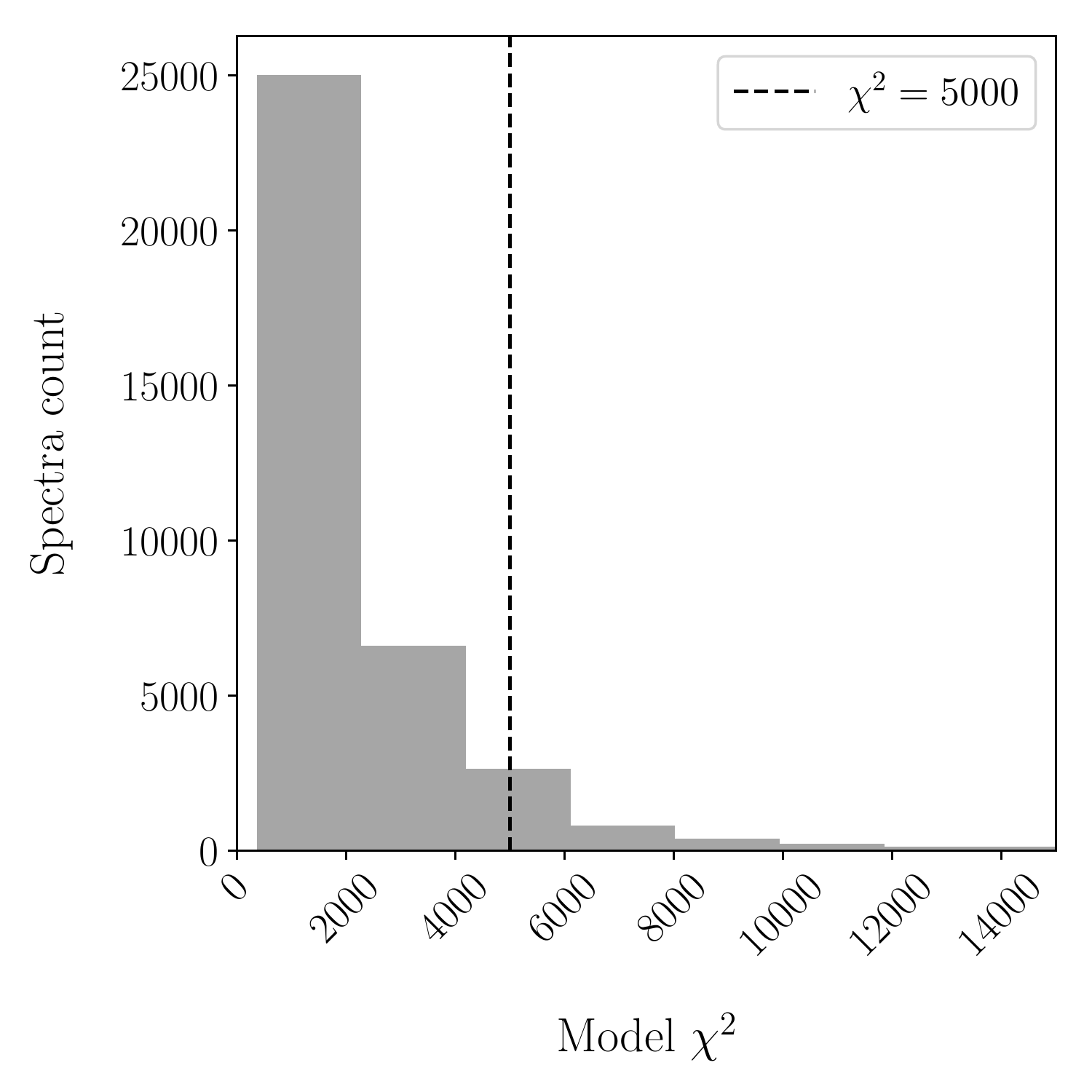}
\caption{Distribution of ${\chi}^2$ values for \cannon model fit for all spectra in the survey set. Since the majority of stars had $\chi^{2}$ $<$ 5,000 (corresponding to $\chi^2$ less than $\sim$3 per degree of freedom), we eliminate any object that had a $\chi^{2}$ greater than this threshold (see Section~\ref{subsec:final_eliminations}). \label{fig:chisq_cutoffs}}
\end{figure}

We visually inspected these failed cases, noting that in many instances the LAMOST spectrum had a very large positive or negative outlier feature, perhaps due to bad pixels that were not masked. However, there were also instances where visually the spectrum did not appear unusual. There were no other obvious trends in the characteristics of the failed spectra; they covered a large \snr\ range and included both early and late subtypes. In conclusion, we were not able to determine an obvious common cause for this failure. In any case, these have been completely removed from our final characterized set in Table~\ref{tab:final_test_results} and what is shown in Figures \ref{fig:chisq_cutoffs}, \ref{fig:teff_lamost_subtype_validation}, and \ref{fig:survey_corner}.

Finally, we also eliminated a small number of stars whose final goodness-of-fit statistic ($\chi^{2}$) from \cannon was very poor. This is shown in Figure~\ref{fig:chisq_cutoffs}, where we adopted a cutoff of $\chi^2 <$ 5,000, corresponding to a $\chi^2$ of $\sim$3 per degree of freedom.

\subsection{Model Parameters} \label{subsec:model_parameters}

The parameters of the M~dwarfs \cannon has calculated from LAMOST DR1 spectra are as expected for their spectral types. Their ranges are \teffcannonrange, \radiuscannonrange, \masscannonrange, and \lumcannonrange. Note that the relatively smaller number of characterized stars at the coolest \teff\ (latest spectral subtypes) is at least in part due to our quality cuts that require \snr\ $> 50$ and that our training set had fewer stars in this range.

\begin{figure}[!ht]
\includegraphics[width=\linewidth]{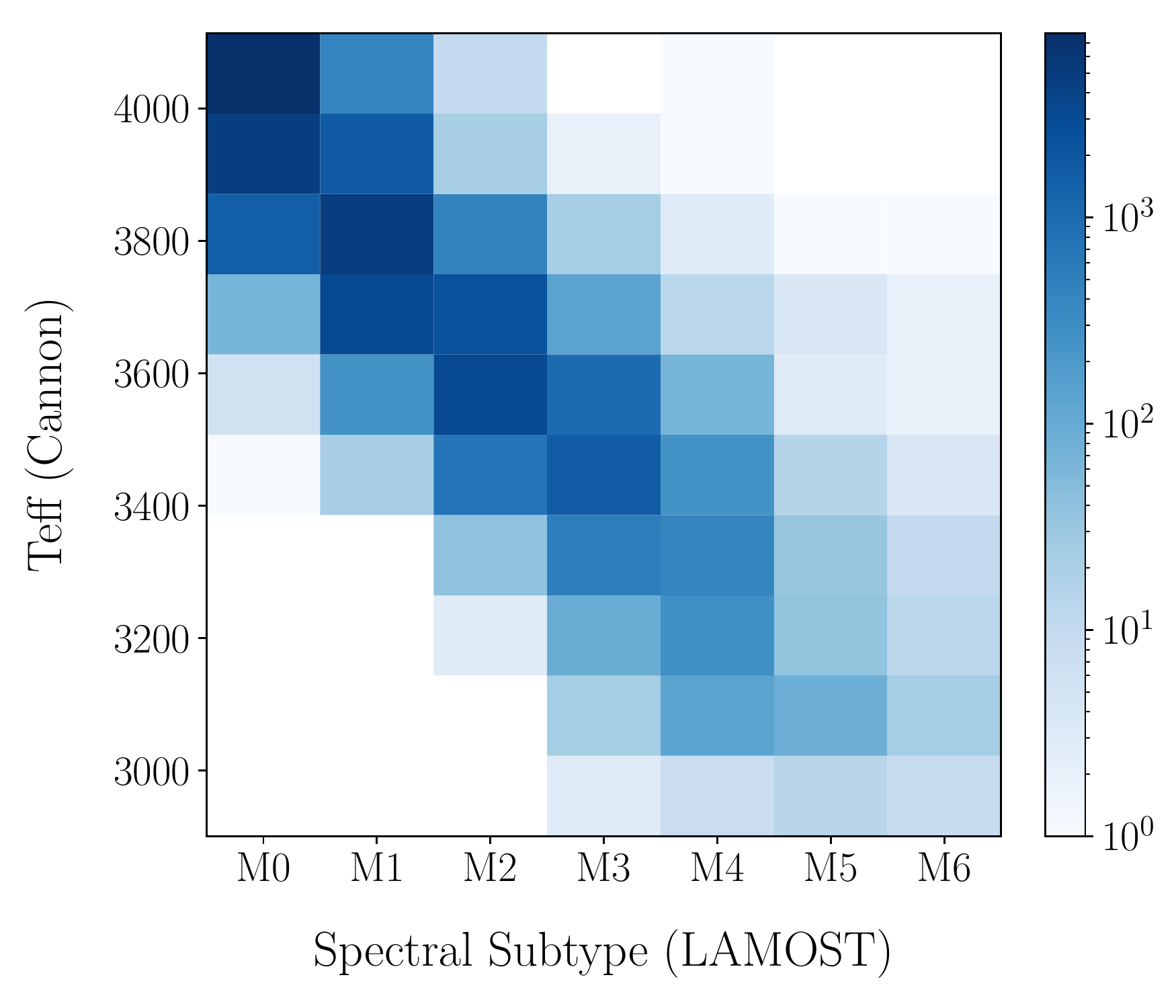}
\caption{Comparison of subtype as reported in the LAMOST DR1 M catalog \citep{Guo:2015} to the outputed model-derived \teff\ by \cannon. 
\label{fig:teff_lamost_subtype_validation}}
\end{figure}

\begin{figure*}[!ht]
\includegraphics[scale=0.7]{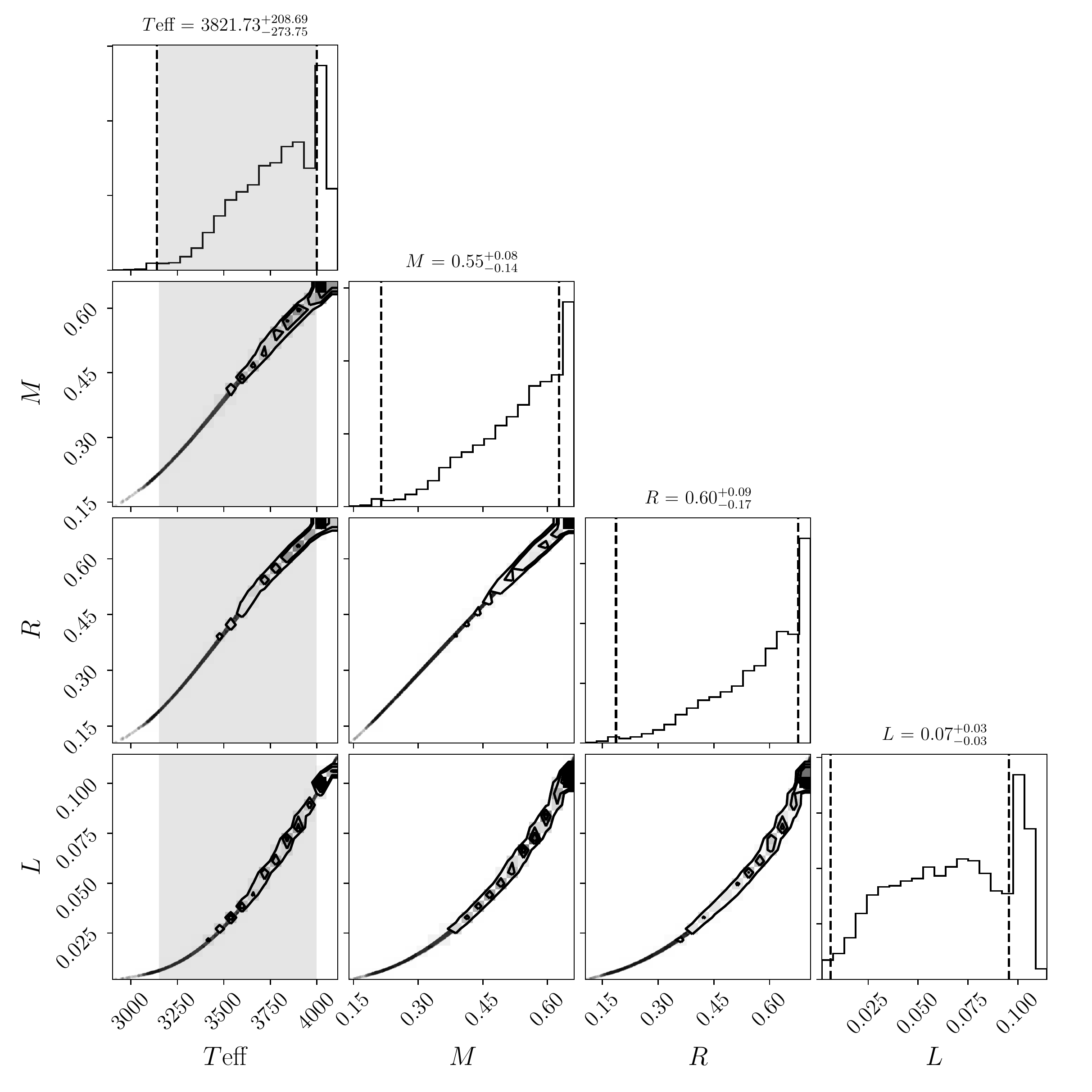}
\caption{The basic parameter space of our characterized survey set as discussed in Section \ref{subsec:model_parameters}. Highlighted in gray is the \teff\ for which the \cannon model we have trained is applicable;
dashed lines represent the other parameter ranges corresponding to this applicable \teff\ range.}
\label{fig:survey_corner}
\end{figure*}

While the LAMOST pipeline does not provide a \teff\ estimate for most M~dwarfs in the catalog, the catalog does provide an estimated subtype,
which we used as a sanity check on the \teff\ output by our trained model (see Figure~\ref{fig:teff_lamost_subtype_validation}). We find that there is a reasonable progression from higher to lower \teff\ derived from \cannon\ corresponding to increasingly later subtypes reported by LAMOST. The overall distribution of model-derived parameters of the newly characterized \testfinalcountdups LAMOST spectra can be seen in Figure~\ref{fig:survey_corner}.

\subsection{Error Estimation} \label{subsec:error_estimation}

As described above, we experimented by running \cannon\ with many different training subsets and assessing the impact of different input variables on the scatter in the resulting output parameters. We found that \snr\ is by far the most important predictor of the parameter scatter (see Figure~\ref{fig:snr_function_error_estimate}) versus other potential factors (e.g., continuum normalization method, different pixel masks, etc). Therefore, we proceed to derive stellar parameter errors for a particular spectrum as a function of its \snr. Figure~\ref{fig:snr_function_error_estimate} shows the empirical relationship between the standard deviation of the residuals of each label from the validation step and the \snr. 

We have opted to approximate the relationship between parameter scatter and \snr\ with a simple polynomial function. This is the simplest form that appears to reasonably well represent the empirical relationship---it is not derived from first principles---however, we believe it fully suffices for our purposes here. Table~\ref{tab:error_coefficents} reports the coefficients of the fitted polynomial relationships, which we use to report the final estimated uncertainties on the labels that we report in Table~\ref{tab:final_test_results}. 

\begin{deluxetable*}{rc|llll} 
\tablecolumns{6}
\tablehead{Estimated stellar parameter error as a function of \snr}
\tablehead{$Y$ & units & \colhead{$a$} & \colhead{$b$} & \colhead{$c$} & \colhead{error floor}}
\startdata
$\sigma$\teff & K & \SI{2.594e-3}{} & \num{-0.942} & \num{174.639} & \num{89.2}\\
$\sigma$\R & \rsun & \num{1.3e-6} & \num{-4.692e-4} & \num{8.688e-2} & \num{0.0432}\\
$\sigma$\M & \msun & \num{1.6e-6} & \num{-5.841e-4} & \num{0.106} & \num{0.0516}\\
$\sigma$\Lum & \lsun & \num{2.0e-7} & \num{-6.3e-5} & \num{1.623e-2} & \num{0.0097}\\
\enddata
\tablecomments{$Y(X) = a + bX + cX^2$, where $X\equiv$ \snr.}
\caption{Polynomial fit of parameter error versus \snr\ (see Figure \ref{fig:snr_function_error_estimate}). 
We 
adopt a minimum error floor for stars with \snr\ $>$ 225, as indicated in the final column (red solid lines in Figure~\ref{fig:snr_function_error_estimate}). 
\label{tab:error_coefficents}}
\end{deluxetable*}

We do note that the uncertainties we infer in this way for the high \snr\ training sample are very similar to those reported by \citet{Muirhead:2018}, even though \cannon\ model does not incorporate label uncertainties in its training. In other words, the model appears to be naturally reproducing the expected errors for the highest quality sample, effectively setting the error floor for the output parameters.

\section{Summary} \label{sec:SUMMARY}
We were successfully able to extend the range of use for \cannon in determining properties of M~dwarfs to low-resolution, low signal-to-noise, optical spectra. We obtained spectra from the LAMOST DR1 M~dwarf catalog and properties from the TESS Cool Dwarf Catalog, and performed a cross-label transfer similar to the procedure detailed in \citet{Ho:2017_labeltransfer}. First, we performed quality checks on the spectra before implementing the model, such as filtering out poor spectra and establishing minimum signal-to-noise ratio (\snr) cutoffs (Section~\ref{subsubsec:spectrum_filtering}, Section~\ref{subsec:snr_quality_cutoffs}). We selected a training subset with the best available (\snr$>$250) LAMOST spectra that also had reasonable TESS Cool Dwarf Catalog parameters to train the model (Section~\ref{subsec:get_TESS_parameters}). We then assessed the model's validity and calculated parameter errors as a function of \snr\ (Section~\ref{subsec:model_validation}, Section~\ref{subsec:error_estimation}). We also eliminated objects from our final characterized survey set according to poor $\chi^{2}$ 
and kept only objects for which all four model parameters (\teff, \R, \M, \Lum) could be determined successfully.

In the end, were able to apply this method to determine the properties of $\sim$30,000 M~dwarfs observed by LAMOST that had not been otherwise characterized. 
We achieve typical uncertainties of \teffsig\ in \teff\ ($\sim$3\%), \radiussig~\rsun\ ($\sim$14\%) in radius, \masssig~\msun\ ($\sim$12\%) in mass, and \lumsig~\lsun\ ($\sim$20\%) in luminosity, driven almost entirely by the precision and range of the training set. The model presented here can be rapidly applied to future LAMOST data releases, significantly extending the samples of well characterized M~dwarfs across the sky using new and exclusively data-based modeling methods.

There is room for future improvements. One particular area for improved accuracy would be to incorporate parameter measurement uncertainties as as an additional weight for the model. \cannon approach can also be sensitive to outliers, especially if the training set is small. 
We therefore recommend careful scrutiny and removal of outliers, 
such as by the procedures we adopted in Sections~\ref{subsubsec:spectrum_filtering} and \ref{subsec:get_TESS_parameters}.

Even so, this work demonstrates that \cannon model can be applied to M~dwarfs with a smaller training set than those used in previous works \citep[e.g.,][]{Ness:2015, Ness:2016, Ho:2017_labeltransfer, Ho:2017_lamostgiants}, and can be used specifically to determine basic stellar properties from LAMOST M~dwarf spectra with high accuracy and speed.

\acknowledgments
B.G.\ acknowledges partial funding support from NSF PAARE grant AST-1358862 through the Fisk-Vanderbilt Masters-to-PhD Bridge Program. B.R-A. acknowledges funding support from FONDECYT through grant 11181295.

\software{The Cannon \citep{Ness:2015}, 
corner \citep{corner}, matplotlib \citep{matplotlib}, numpy \citep{numpy}}

\bibliographystyle{apj}
\bibliography{citations.bib}

\begin{thebibliography}{}
\expandafter\ifx\csname natexlab\endcsname\relax\def\natexlab#1{#1}\fi

\bibitem[{{Allard} {et~al.}(2011){Allard}, {Homeier}, \&
  {Freytag}}]{Allard:2011}
{Allard}, F., {Homeier}, D., \& {Freytag}, B. 2011, in Astronomical Society of
  the Pacific Conference Series, Vol. 448, 16th Cambridge Workshop on Cool
  Stars, Stellar Systems, and the Sun, ed. C.~{Johns-Krull}, M.~K. {Browning},
  \& A.~A. {West}, 91

\bibitem[{{Behmard} {et~al.}(2019){Behmard}, {Petigura}, \&
  {Howard}}]{Behmard:2019}
{Behmard}, A., {Petigura}, E.~A., \& {Howard}, A.~W. 2019, arXiv e-prints,
  arXiv:1904.00094

\bibitem[{Foreman-Mackey(2016)}]{corner}
Foreman-Mackey, D. 2016, The Journal of Open Source Software, 24,
  doi:10.21105/joss.00024

\bibitem[{Guo {et~al.}(2015)Guo, Yi, Luo, Wang, Bai, Yang, Song, Chen, Chen,
  Zuo, Du, Zhang, Li, Kong, Wang, Wu, Wu, Zhao, Zhang, Hou, Wang, \&
  Yang}]{Guo:2015}
Guo, Y.-X., Yi, Z.-P., Luo, A.-L., {et~al.} 2015, Research in Astronomy and
  Astrophysics, 15, 1182

\bibitem[{{Ho} {et~al.}(2017{\natexlab{a}}){Ho}, {Rix}, {Ness}, {Hogg}, {Liu},
  \& {Ting}}]{Ho:2017_lamostgiants}
{Ho}, A. Y.~Q., {Rix}, H.-W., {Ness}, M.~K., {et~al.} 2017{\natexlab{a}}, \apj,
  841, 40

\bibitem[{{Ho} {et~al.}(2017{\natexlab{b}}){Ho}, {Ness}, {Hogg}, {Rix}, {Liu},
  {Yang}, {Zhang}, {Hou}, \& {Wang}}]{Ho:2017_labeltransfer}
{Ho}, A. Y.~Q., {Ness}, M.~K., {Hogg}, D.~W., {et~al.} 2017{\natexlab{b}},
  \apj, 836, 5

\bibitem[{Hunter(2007)}]{matplotlib}
Hunter, J.~D. 2007, Computing in Science \& Engineering, 9, 90

\bibitem[{{Husser} {et~al.}(2013){Husser}, {Wende-von Berg}, {Dreizler},
  {Homeier}, {Reiners}, {Barman}, \& {Hauschildt}}]{Husser:2013}
{Husser}, T.~O., {Wende-von Berg}, S., {Dreizler}, S., {et~al.} 2013, \aap,
  553, A6

\bibitem[{{L{\'e}pine} {et~al.}(2013){L{\'e}pine}, {Hilton}, {Mann}, {Wilde},
  {Rojas-Ayala}, {Cruz}, \& {Gaidos}}]{Lepine:2013}
{L{\'e}pine}, S., {Hilton}, E.~J., {Mann}, A.~W., {et~al.} 2013, \aj, 145, 102

\bibitem[{{Luo} {et~al.}(2015){Luo}, {Zhao}, {Zhao}, {Deng}, {Liu}, {Jing},
  {Wang}, {Zhang}, {Shi}, {Cui}, {Chu}, {Li}, {Bai}, {Cai}, {Cao}, {Cao},
  {Carlin}, {Chen}, {Chen}, {Chen}, {Chen}, {Chen}, {Chen}, {Chen},
  {Christlieb}, {Chu}, {Cui}, {Dong}, {Du}, {Fan}, {Feng}, {Fu}, {Gao}, {Gong},
  {Gu}, {Guo}, {Han}, {He}, {Hou}, {Hou}, {Hou}, {Hu}, {Hu}, {Hu}, {Huo},
  {Jia}, {Jiang}, {Jiang}, {Jiang}, {Jin}, {Kong}, {Kong}, {Lei}, {Li}, {Li},
  {Li}, {Li}, {Li}, {Li}, {Li}, {Li}, {Li}, {Li}, {Li}, {Li}, {Liang}, {Lin},
  {Liu}, {Liu}, {Liu}, {Liu}, {Lu}, {Luo}, {Mao}, {Newberg}, {Ni}, {Qi}, {Qi},
  {Shen}, {Shi}, {Song}, {Song}, {Su}, {Su}, {Tang}, {Tao}, {Tian}, {Wang},
  {Wang}, {Wang}, {Wang}, {Wang}, {Wang}, {Wang}, {Wang}, {Wang}, {Wang},
  {Wang}, {Wang}, {Wang}, {Wang}, {Wang}, {Wang}, {Wang}, {Wang}, {Wang},
  {Wang}, {Wei}, {Wei}, {Wu}, {Wu}, {Wu}, {Wu}, {Wu}, {Xing}, {Xu}, {Xu}, {Xu},
  {Yan}, {Yang}, {Yang}, {Yang}, {Yang}, {Yao}, {Yu}, {Yuan}, {Yuan}, {Yuan},
  {Yuan}, {Zhai}, {Zhang}, {Zhang}, {Zhang}, {Zhang}, {Zhang}, {Zhang},
  {Zhang}, {Zhang}, {Zhao}, {Zhou}, {Zhou}, {Zhu}, {Zhu}, {Zou}, \&
  {Zuo}}]{Luo:2015}
{Luo}, A.~L., {Zhao}, Y.~H., {Zhao}, G., {et~al.} 2015, arXiv e-prints,
  arXiv:1505.01570

\bibitem[{{Mann} {et~al.}(2015){Mann}, {Feiden}, {Gaidos}, {Boyajian}, \& {von
  Braun}}]{Mann:2015}
{Mann}, A.~W., {Feiden}, G.~A., {Gaidos}, E., {Boyajian}, T., \& {von Braun},
  K. 2015, \apj, 804, 64

\bibitem[{{Mann} {et~al.}(2019){Mann}, {Dupuy}, {Kraus}, {Gaidos}, {Ansdell},
  {Ireland}, {Rizzuto}, {Hung}, {Dittmann}, {Factor}, {Feiden}, {Martinez},
  {Ru{\'\i}z-Rodr{\'\i}guez}, \& {Chia Thao}}]{Mann:2019}
{Mann}, A.~W., {Dupuy}, T., {Kraus}, A.~L., {et~al.} 2019, \apj, 871, 63

\bibitem[{{Muirhead} {et~al.}(2018){Muirhead}, {Dressing}, {Mann},
  {Rojas-Ayala}, {L{\'e}pine}, {Paegert}, {De Lee}, \&
  {Oelkers}}]{Muirhead:2018}
{Muirhead}, P.~S., {Dressing}, C.~D., {Mann}, A.~W., {et~al.} 2018, \aj, 155,
  180

\bibitem[{{Ness} {et~al.}(2015){Ness}, {Hogg}, {Rix}, {Ho}, \&
  {Zasowski}}]{Ness:2015}
{Ness}, M., {Hogg}, D.~W., {Rix}, H.~W., {Ho}, A. Y.~Q., \& {Zasowski}, G.
  2015, \apj, 808, 16

\bibitem[{{Ness} {et~al.}(2016){Ness}, {Hogg}, {Rix}, {Martig}, {Pinsonneault},
  \& {Ho}}]{Ness:2016}
{Ness}, M., {Hogg}, D.~W., {Rix}, H.~W., {et~al.} 2016, \apj, 823, 114

\bibitem[{Oliphant(2006)}]{numpy}
Oliphant, T.~E. 2006, A guide to NumPy, Vol.~1 (Trelgol Publishing USA)

\bibitem[{{Rojas-Ayala} {et~al.}(2012){Rojas-Ayala}, {Covey}, {Muirhead}, \&
  {Lloyd}}]{Rojas-Ayala:2011}
{Rojas-Ayala}, B., {Covey}, K.~R., {Muirhead}, P.~S., \& {Lloyd}, J.~P. 2012,
  \apj, 748, 93

\bibitem[{{Shields} {et~al.}(2016){Shields}, {Ballard}, \&
  {Johnson}}]{Shields:2016}
{Shields}, A.~L., {Ballard}, S., \& {Johnson}, J.~A. 2016, \physrep, 663, 1

\end{thebibliography}
\end{document}